\documentclass[traditabstract,letter]{aa}

\pdfoutput=1
\usepackage{amsmath}
\usepackage{graphicx}
\usepackage{txfonts}
\usepackage{natbib}
\usepackage{longtable,lscape}
\usepackage[plainpages=False,bookmarks=False,colorlinks=True, citecolor=blue,linkcolor=blue]{hyperref}
\usepackage{ulem}
\usepackage{comment}
\usepackage{cases}

\newcommand{\Nsn}{76}
\newcommand{\snia}{SN~Ia}
\newcommand{\sneia}{SNe~Ia}
\newcommand{\SiIIawlength}{4131}
\newcommand{\Sia}{Si~\textsc{ii} $\lambda$\SiIIawlength}
\newcommand{\Sic}{Si~\textsc{ii} $\lambda$6355}
\newcommand{\CaHK}{Ca~\textsc{ii}~H\&K}
\newcommand{\EWSi}{\ensuremath{\textsc{ew}^{\mathrm{Si}}}}
\newcommand{\EWCa}{\ensuremath{\textsc{ew}^{\mathrm{Ca}}}}
\newcommand{\Rv}{\ensuremath{R_V}}
\newcommand{\dmz}{\ensuremath{\Delta m_{15}}}

\begin{document}
\date{\today}
\title{The reddening law of Type Ia Supernovae: separating intrinsic
  variability from dust using equivalent widths}

\titlerunning{Reddening law of SNe~Ia} \authorrunning{N. Chotard et
  al. (Nearby Supernova Factory)}

\author
{
  The Nearby Supernova Factory: \\
  N.~Chotard\inst{1},  
  E.~Gangler\inst{1},  
  G.~Aldering\inst{2},  
  P.~Antilogus\inst{3},  
  C.~Aragon\inst{2},  
  S.~Bailey\inst{2},  
  C.~Baltay\inst{4},  
  S.~Bongard\inst{3},  
  C.~Buton\inst{5},  
  A.~Canto\inst{3},  
  M.~Childress\inst{2,6},  
  Y.~Copin\inst{1},  
  H.~K.~Fakhouri\inst{2,6},  
  E.~Y.~Hsiao\inst{2},  
  M.~Kerschhaggl\inst{5},  
  M.~Kowalski\inst{5},  
  S.~Loken\inst{2},  
  P.~Nugent\inst{7,8},  
  K.~Paech\inst{5},  
  R.~Pain\inst{3},  
  E.~Pecontal\inst{9},  
  R.~Pereira\inst{1},  
  S.~Perlmutter\inst{2,6},  
  D.~Rabinowitz\inst{4},  
  K.~Runge\inst{2},  
  R.~Scalzo\inst{4,10},  
  G.~Smadja\inst{1},  
  C.~Tao\inst{11,12},  
  R.~C.~Thomas\inst{7},  
  B.~A.~Weaver\inst{13},
  and C.~Wu\inst{3,14}  
}

\institute{Universit\'e de Lyon, F-69622, France ; 
Universit\'e de Lyon 1, Villeurbanne ;
CNRS/IN2P3, Institut de Physique Nucl\'eaire de Lyon
\and
Physics Division, Lawrence Berkeley National
Laboratory, 1 Cyclotron Road, Berkeley, CA 94720
\and
LPNHE,
Universit\'e Pierre et Marie Curie Paris 6, Universit\'e Paris
Diderot Paris 7, CNRS-IN2P3, 75252 Paris Cedex 05,
France
\and
Department of Physics, Yale University, New Haven, CT 06250-8121
\and
Physikalisches Institut Universitat Bonn, Nussallee 12 53115
Bonn, Germany
\and
Department of Physics, University of California Berkeley, 
366 LeConte Hall MC 7300, Berkeley, CA, 94720-7300
\and
Computational Cosmology Center, 
Lawrence Berkeley National Laboratory, 
1 Cyclotron Road,
Berkeley, CA 94720, USA
\and
Department of Astronomy, University of California, Berkeley, CA 94720-3411, USA
\and
Observatoire de Lyon, Saint-Genis Laval, F-69230, Universit\'e de
Lyon, Lyon, F-69003, France ; Universit\'e Lyon 1
\and
Australian National University, Mt. Stromlo Observatory, The RSAA, Weston Creek, ACT 2611 Australia.
\and
Centre de Physique des Particules de Marseille, 163, avenue de Luminy - Case 902 - 13288 Marseille Cedex 09, France
\and
Tsinghua Center for Astrophysics, Tsinghua University, Beijing 100084, China 
\and
New York University, Center for Cosmology and Particle Physics, 4
Washington Place, New York, NY 10003
\and
National Astronomical Observatories, Chinese Academy of Sciences,
Beijing 100012, China
}

\begin{abstract}{ We employ 76 type Ia supernovae with optical
    spectrophotometry within 2.5 days of B-band maximum light obtained
    by the Nearby Supernova Factory to derive the impact of Si and Ca
    features on supernovae intrinsic luminosity and determine a dust
    reddening law. We use the equivalent width of \Sia\ in place of
    light curve stretch to account for first-order intrinsic
    luminosity variability. The resultant empirical spectral reddening
    law exhibits strong features associated with Ca~\textsc{ii} and
    \Sic. After applying a correction based on the \CaHK\ equivalent
    width we find a reddening law consistent with a Cardelli
    extinction law.  Using the same input data, we compare this result
    to synthetic rest-frame UBVRI-like photometry in order to mimic
    literature observations.  After corrections for signatures
    correlated with \Sia\ and \CaHK\ equivalent widths, and
    introducing an empirical correlation between colors, we determine
    the dust component in each band. We find a value of the
    total-to-selective extinction ratio, $\Rv=2.8\pm0.3$. This agrees
    with the Milky~Way value, in contrast to the low \Rv\ values found
    in most previous analyses. This result suggests that the
    long-standing controversy in interpreting SN~Ia colors and their
    compatibility with a classical extinction law, critical to their
    use as cosmological probes, can be explained by the treatment of
    the dispersion in colors, and by the variability of features
    apparent in SN~Ia spectra.}

\end{abstract}


\maketitle


\section{Introduction}

Type Ia supernovae (\sneia) luminosity distances are measured via the
standardization of their light curves using brightness-width (stretch,
$x_1$, \dmz) and color corrections \citep{Phillips1993, Tripp98,
  Guy2007, Jha2007}.  While the determination of the intrinsic
dispersion related to light curve shape is subject to small
differences between fitters, the manner in which color is linked to
dust is still controversial, as it may be affected by additional yet
unidentified intrinsic dispersion.  Whereas earlier work used the
total-to-selective extinction ratio of the Milky~Way, \Rv=3.1, direct
estimates from supernovae (SNe) Hubble diagram fits lead to lower
values, from \Rv=1.7 to \Rv=2.5 \citep{Hicken09a,Tripp98,Wang09b}.
While the derivation of this value is subject to assumptions about the
natural color dispersion of SNe \citep{Freedman09,Guy10}, the reason
for a difference between SNe data and the Milky~Way average result had
remained unknown.
 
Equivalent widths are good spectral indicators for addressing this
issue as they probe the intrinsic variability of \sneia\, and by
construction they have little dependence on extinction due to their
narrow wavelength baseline.  \cite{Arsenijevic2008} showed a strong
correlation between the equivalent width of the \Sia\ feature and the
SALT2 $x_1$ width parameter \citep{Guy2007}, and \cite{Bronder2008}
showed its correlation with $M_B$.  \cite{Walker2010} used it to
standardize the Hubble diagram, but were hindered from drawing firm
conclusions by the quality of the low redshift data.

In this work, we take advantage of the Nearby Supernova Factory
(SNfactory) spectrophotometric sample to revisit these conclusions,
using both spectral data and derived UBVRI-like synthetic
photometry. We present in Section~\ref{data} the \sneia\ sample and
the definition of the \Sia\ and \CaHK\ equivalent widths, which will
be used in Section~\ref{intrcorr} to correct the Hubble
residuals. These corrected magnitudes are used to derive the relative
absorption in each wavelength band, $\delta A_{\lambda}$. The
correlations between the $\delta A_{\lambda}$ from SN to SN across
different bands provide the reddening law, as described in
Section~\ref{model}, as well as the dispersion matrix between
bands. In Section~\ref{results} it is shown that the resulting
reddening law agrees with a Cardelli extinction law
\citep[CCM,][]{Cardelli1989,ODonnell94}. Our reddening law has a value
of \Rv\ that agrees with the Milky-Way value of $3.1$, when the proper
dispersion matrix is used. We then discuss these results in
Section~\ref{discussion} and conclude in Section~\ref{conclusion}.


\section{Data-set and derived quantities}
\label{data}

This analysis uses flux calibrated spectra of \Nsn\ \sneia\ obtained
by the SNfactory collaboration with its SNIFS instrument
\citep{Aldering2002} on the University of Hawaii 2.2-meter telescope
on Mauna Kea.  This subset is selected in the same way as in
\cite{bailey2009}, using only SNe having a measured spectrum within
$2.5$ days of B-band maximum, but with an enlarged data-set with a
redshift range of $0.007<z<0.09$. The SALT2 $x_{1}$ and $c$ parameters
(Fig.\ref{EWSi4000_vs_mag}), as well as the spectra phases with
respect to the B-band maximum light, are derived using fits of the
full light curves in three observer frame top hat bandpasses
corresponding approximately to BVR.

After flux calibration, host galaxy subtraction and correction for
Milky~Way extinction, the \Nsn\ spectra are transformed into rest
frame. For the spectral analysis, they are rebinned with a resolution
of $1500~\textnormal{km}~\textnormal{s}^{-1}$.  In addition, synthetic
magnitudes are derived from the spectra after photon integration in a
set of five UBVRI-like top-hat bandpasses with a constant resolution
over the whole spectral range ($3276-8635$~\AA, see
Fig.\ref{cardellifit} and caption).  The uncorrected Hubble residuals
$\delta M(\lambda)=M(\lambda)-\langle M(\lambda)\rangle$ are
independently computed for each band of mean wavelength $\lambda$,
relative to a flat $\Lambda$CDM universe with $\Omega_{\rm M}=0.28$
and
$H_{0}=70~\textnormal{km}~\textnormal{s}^{-1}~\textnormal{Mpc}^{-1}$.
For the spectral analysis, the Hubble residuals are corrected for
phase dependence using linear interpolation.  For broad bands, we
instead interpolate the magnitudes to the date of B maximum using the
light curve shape of each \snia\ as defined by the SALT2 $x_{1}$
parameter, but with each band's peak magnitude fitted separately. The
data are presented in the online Table~\ref{datatable}. Error on
Hubble residuals include statistical, calibration ($\sim 2\%$ for peak
luminosity) and redshift uncertainties. They are correlated between
bands with a correlation coefficient varying from $0.64$ to $0.98$.

The equivalent widths \EWSi\ and \EWCa\ corresponding to the \Sia\ and
the \CaHK\ features are computed in the same way as in
\citet[eq~1]{Bronder2008}. The errors are derived using a Monte Carlo
procedure taking into account photon noise and the impact of the
method used to select the feature boundaries.  \EWSi\ and \EWCa\ are
insensitive to extinction by dust, changing by less than 1\% when
adding an artificial reddening with a CCM law with $\Rv=3.1$ and
$E(B-V)=0.5$.


\section{Method}
\label{method}

\subsection{Intrinsic corrections and absorption measurements}
\label{intrcorr}

\begin{figure}
  \centering
  \includegraphics[width=0.49\textwidth]{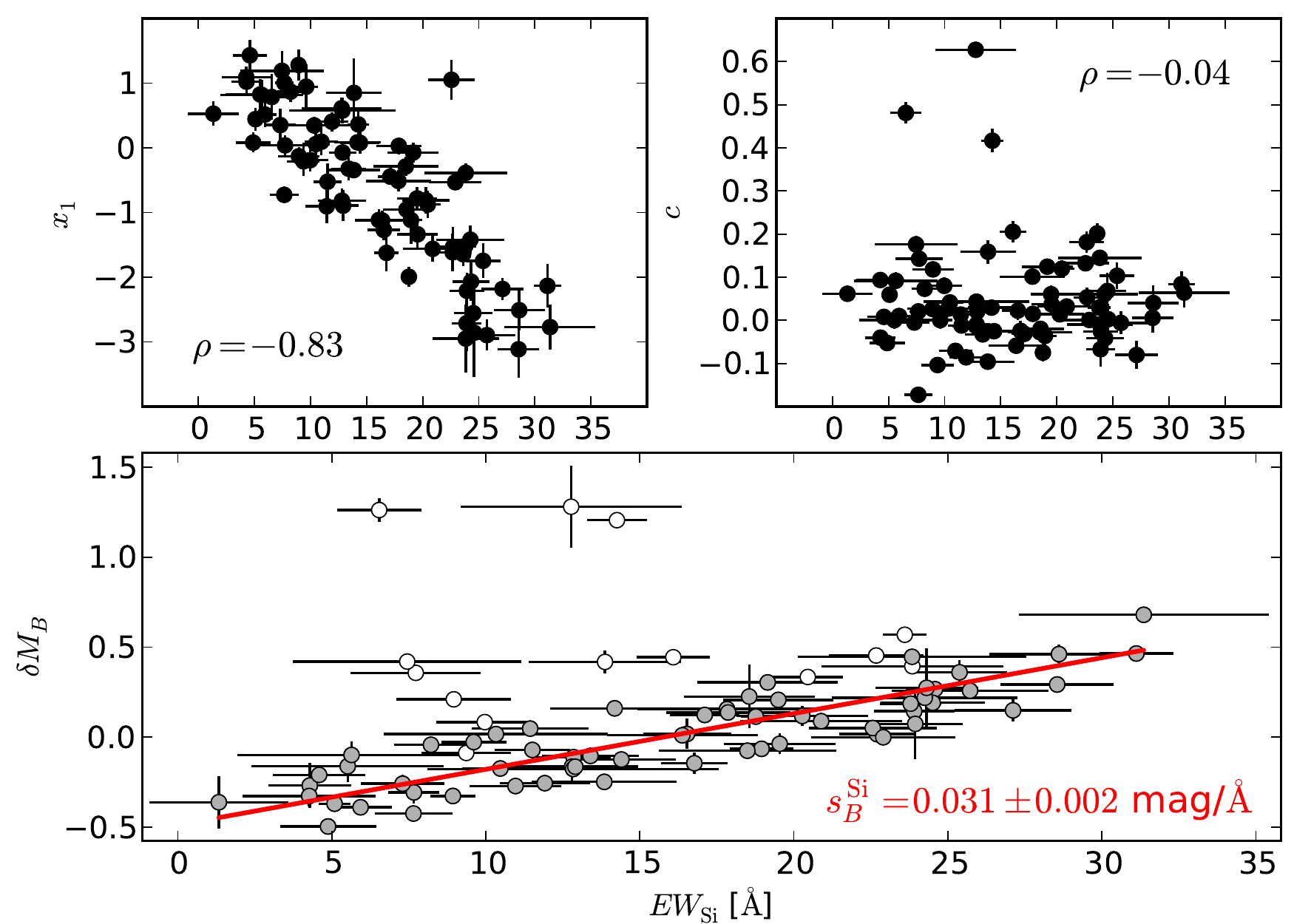}
  \caption{\footnotesize{Correlations of \EWSi, SALT2 $x_{1}$
      (\textit{top left}) and $c$ (\textit{top right}) parameters and
      measured peak absolute magnitude up to a constant term, $\delta
      M_B$ (\textit{bottom}).  The \textit{open circles} in the bottom
      panel are the data points excluded from the fit, which is
      displayed as a \textit{solid line}. $\vec{s_{B}^{\mathrm{Si}}}$ is
      equivalent for the B band to the red curve shown in
      Fig.~\ref{cardellifit}\textcolor{blue}{a}.}}
  \label{EWSi4000_vs_mag}
\end{figure}

The Hubble residuals exhibit a dependency on observables such as
\EWSi\ and \EWCa\ which are uncorrelated with dust extinction. As
shown in Fig.~\ref{EWSi4000_vs_mag} as an example, the $\delta M_B$
dependence on \EWSi\ exhibits a linear behaviour, with an asymmetrical
magnitude dispersion attributed to extinction and remaining intrinsic
variability. Similar dependence with equivalent widths is found for
other bands.  We may thus model the Hubble residuals for a given SN,
$i$, as a sum of intrinsic and dust components, making various
assumptions about the number of spectral energy distribution (SED)
correction vectors, $\vec{s_\lambda}$, from none (Eq.\ref{eq:0}) to
two (Eq.\ref{eq:2}):
\begin{subnumcases}{\delta M_{\lambda,i}=} 
  \delta A^{0}_{\lambda,i} \label{eq:0} \\
  \textsc{ew}^\mathrm{Si}_{i} \vec{s^{\mathrm{Si}}_\lambda}  + \delta
  A_{\lambda,i}^{\mathrm{Si}}  \label{eq:1} \\
  \textsc{ew}^\mathrm{Si}_{i} \vec{s^{\mathrm{Si}}_\lambda}
  +  \textsc{ew}^\mathrm{Ca}_{i} \vec{s^{\mathrm{Ca}}_\lambda}
  + \delta A_{\lambda,i}^{\mathrm{Si+Ca}} \label{eq:2} 
\end{subnumcases}
After finding $\vec{s^{\mathrm{Si}}_\lambda}$ and the $\delta
A_{\lambda,i}^{\mathrm{Si}}$ by a least squares minimization over all
SNe from Eq.~\ref{eq:1}, with an asymmetrical $2\sigma$ clipping,
$\vec{s^{\mathrm{Si}}_\lambda}$ is kept constant and
$\vec{s^{\mathrm{Ca}}_\lambda}$ and the $\delta
A_{\lambda,i}^{\mathrm{Si+Ca}}$ are then computed in the same
manner. An example of the fit using \EWSi\ for the B band is given is
Fig.~\ref{EWSi4000_vs_mag} (\textit{solid line}), $s^{\mathrm{Si}}_B$
being the correction applied to find the $\delta
A_{B,i}^{\mathrm{Si}}$.


\subsection{The empirical reddening law}
\label{model}

\begin{figure}
  \centering
  \includegraphics[width=0.49\textwidth]{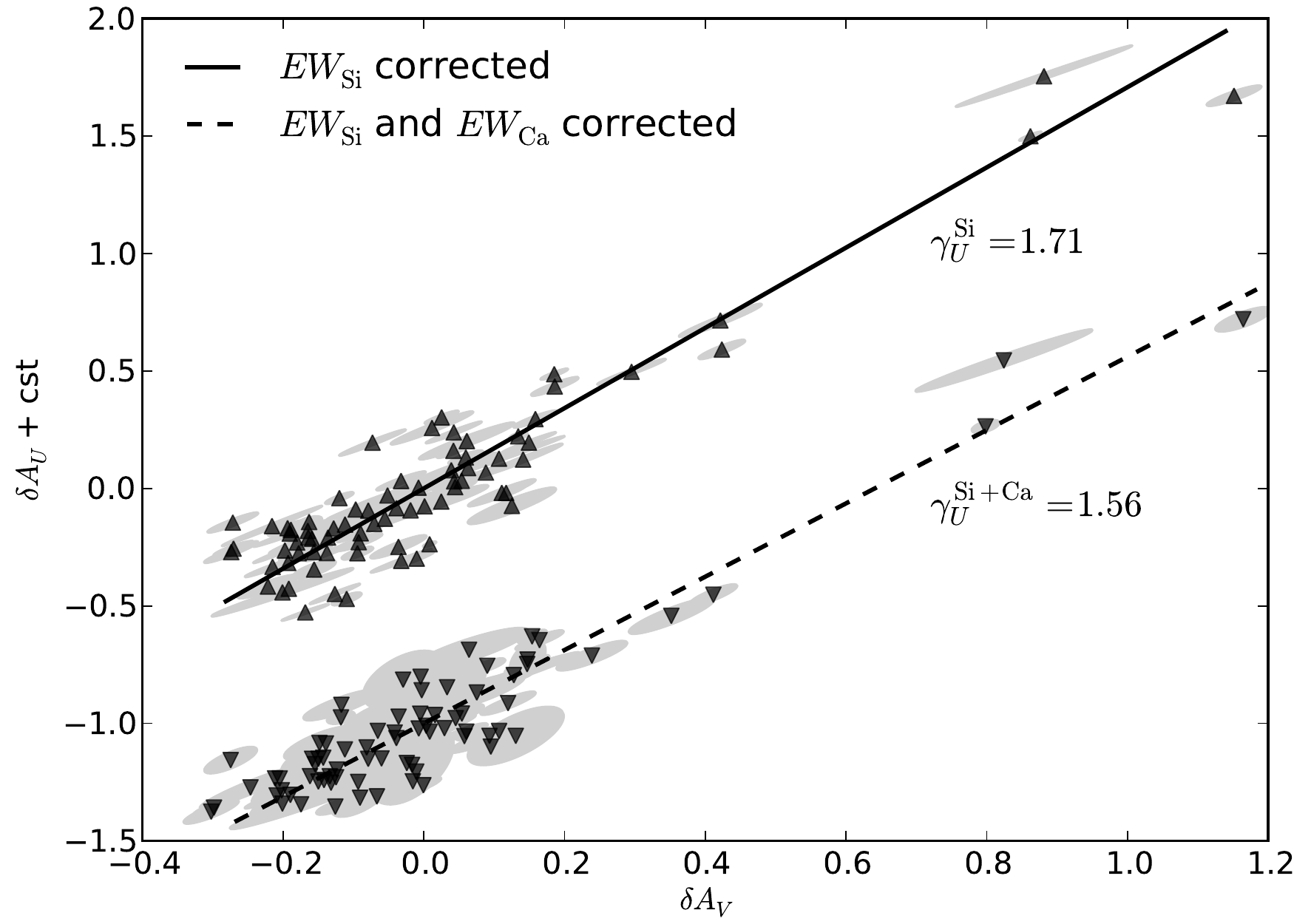}
  \caption{\footnotesize{Relation between $\delta A_U$ and $\delta
      A_V$ after \EWSi\ correction (\textit{triangles up}) and after
      \EWSi\ and \EWCa\ corrections (\textit{triangles down}).  The
      $\delta A_U$ are displayed with an added arbitrary constant. The
      ellipses represent the full measured covariance matrix between
      the two bands. They are enlarged by the additional subtraction
      error after \EWCa\ correction. The result of the fits are also
      displayed.  The $\gamma_{U}^{\mathrm{Si}}$ and
      $\gamma_{U}^{\mathrm{Si+Ca}}$ respectively correspond to the
      ones in Fig.~\ref{cardellifit}\textcolor{blue}{b} and
      Fig.~\ref{cardellifit}\textcolor{blue}{c} for the U band.}}
  \label{Rv}
\end{figure}

As we expect the relation between the $\delta A_\lambda$ to be linear
for dust extinction, as shown in Fig.~\ref{Rv} for the U and V bands,
we model the empirical reddening law as:
\begin{eqnarray}
\delta A_{\lambda,i} = \gamma_\lambda \; \delta A_{V,i}^*  +
\eta_\lambda
\label{eq:model}
\end{eqnarray}
where $\delta A_{\lambda,i}$ are the measured values, the slopes
$\gamma_\lambda$ ($=\partial A_{\lambda}/\partial A_V^*$) the
reddening law coefficients, $\delta A_{V,i}^*$ the fitted relative
extinction for the SN $i$, and $\eta_\lambda$ a free zero point.
$\delta A_{V,i}^*$, $\gamma_\lambda$ and $\eta_\lambda$ are obtained
by a $\chi^2$ fit using the full wavelength covariance matrix $C_i$ of
the $\delta A_{\lambda,i}$ measurements.  However, the data are more
dispersed than their measurement error (Fig.~\ref{Rv}), and another
source of dispersion must be introduced.  We adopt an iterative
approach by using the fit residuals, \mbox{$r_{\lambda,i}=\delta
  A_{\lambda,i} - ( \gamma_\lambda \; \delta A_{V,i}^* + \eta_\lambda
  )$}, to determine the covariance remaining after accounting for the
measurement error covariance, $C_{\lambda_1\lambda_2,i}$. This
empirical covariance matrix, D, is given by:
\begin{eqnarray}
  D_{\lambda_1\lambda_2}=\frac{1}{N}\sum_{i=1}^{N}\
    \left(r_{\lambda_1,i}\,r_{\lambda_2,i} - C_{\lambda_1\lambda_2,i} \right)\ \ \ \ \ 
\label{eq:matrix}
\end{eqnarray}
where $N$ is the number of SNe. In the next iteration, the total
covariance matrix is given by $C_{i}+D$. We have checked that the
converged matrix does not depend on initial conditions.


\section{Results} 
\label{results}

\begin{figure}
  \centering
  \includegraphics[width=0.49\textwidth]{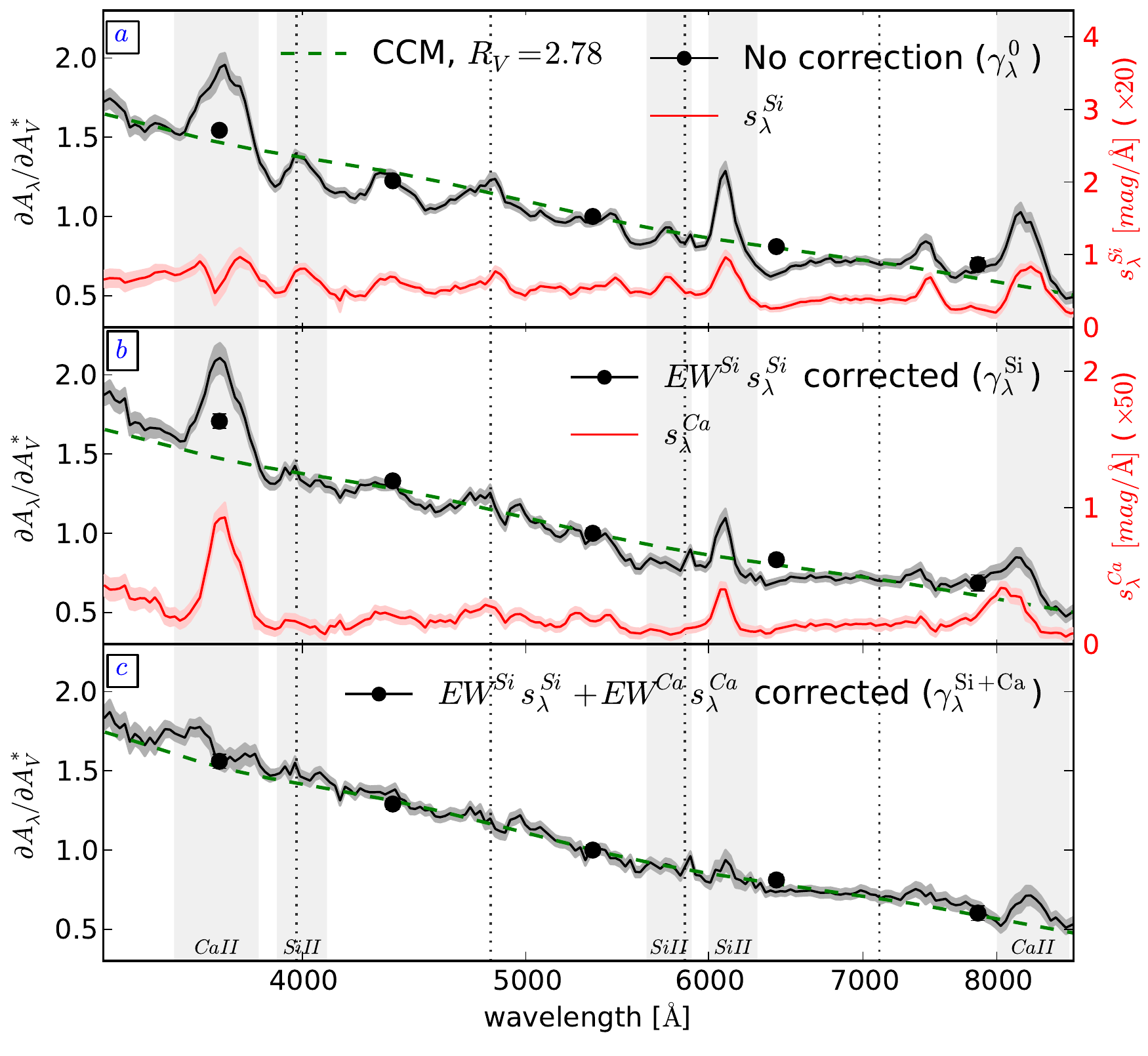}
  \caption{\footnotesize{\textit{Black:}~Reddening law presented as
      $\gamma_{\lambda}\equiv \partial A_{\lambda}/\partial A_V^{*}$
      as a function of wavelength. Filled circles correspond to the
      results obtained using the UBVRI-like bands, curves are for the
      spectral analysis. \textit{Red:}~Linear slope $\vec{s_\lambda}$
      (mag/\AA) of equivalent widths versus $\delta
      M_{\lambda}$. \textit{Dotted lines}:~CCM law fit corresponding
      to the broad bands analysis. \textit{Panel
        \textcolor{blue}{a}:}~$\delta M_{\lambda}$ corrected only for
      the phase dependence (eq.~\ref{eq:0}). \textit{Panel
        \textcolor{blue}{b}:}~$\delta M_{\lambda}$ corrected for phase
      and \EWSi\ (eq.~\ref{eq:1}). \textit{Panel
        \textcolor{blue}{c}:}~$\delta M_{\lambda}$ corrected for
      phase, \EWSi\ and \EWCa\ (eq.~\ref{eq:2}). The vertical dotted
      lines represent the UBVRI-like bands boundaries. The shaded
      vertical bands represent the Si and Ca domain. The shaded band
      around the curves are the statistical errors.}}
  \label{cardellifit}
\end{figure}


\subsection{\EWSi\ and \EWCa\ impacts on the derived extinction law}

Results for the SED correction vector, $\vec{s_\lambda}$, and the
reddening law, $\gamma_\lambda$, are presented in
Fig.~\ref{cardellifit} for different assumptions about the number of
intrinsic components. If SNe were perfect standard candles affected
only by dust as assumed by Eq.~\ref{eq:0} and
Fig.~\ref{cardellifit}\textcolor{blue}{a}, the empirical reddening law
$\gamma_\lambda^{0}$ would be a CCM-like law with an average \Rv\ for
our galaxy sample. However, $\gamma_\lambda^{0}$ clearly exhibits
small-scale SN-like features. These features correlate strongly with
some of the features in the \EWSi\ correction spectrum,
$\vec{s_\lambda^{\mathrm{Si}}}$, derived via Eq.~\ref{eq:1} and
illustrated in red in Fig.~\ref{cardellifit}\textcolor{blue}{a}.

The reddening law obtained after \EWSi\ correction
(Fig.~\ref{cardellifit}\textcolor{blue}{b}) is closer to a CCM law,
except in the \CaHK\ and IR triplet, and the \Sic\ region. This
indicates the presence of a second source of intrinsic variability.
We select \EWCa\ to trace a second spectral correction vector,
$\vec{s_\lambda^{\mathrm{Ca}}}$ (Eq.~\ref{eq:2}), since Ca is clearly
a major contributor to the observed variability and \EWCa\ and \EWSi\
also happen to be uncorrelated ($\rho=0.06\pm0.12$). As shown in
Fig.~\ref{cardellifit}\textcolor{blue}{b},
$\vec{s_\lambda^{\mathrm{Ca}}}$ does a good job of reproducing the
shape of the deviation of $\gamma_\lambda^{Si}$ relative to the CCM
law.

The mean reddening law, $\gamma_\lambda^{Si+Ca}$, obtained after the
additional correction by \EWCa\
(Fig.~\ref{cardellifit}\textcolor{blue}{c}) is a much smoother curve
with small residual features, and agrees well with a CCM extinction
law. Thus it appears that these two components can account for \snia\
spectral variations at optical wavelengths. Any intrinsic component
that might remain would have to be largely uncorrelated with SN
spectral features fixed in wavelength, as well as being coincidentally
compatible with a CCM law.

\subsection{\Rv\ determination}

We apply the same treatment as above to our UBVRI-like synthetic
photometric bands, and find agreement with the spectral analysis, as
shown by the black points in the three panels in
Fig.~\ref{cardellifit}. After the \EWSi\ correction (Eq.~\ref{eq:1},
Fig.\ref{cardellifit}\textcolor{blue}{b}), the U and I band values
deviate significantly from a CCM law, which is recovered after the
full \EWSi\ and \EWCa\ correction (Eq.~\ref{eq:2},
Fig.\ref{cardellifit}\textcolor{blue}{c}).  The empirical fit
presented in subsection~\ref{model} can be forced to follow a CCM
extinction law by substituting \mbox{$ \delta A_{\lambda,i} =
  \left(a_{\lambda} + b_{\lambda}/\Rv \right) \delta A_{V,i}^* +
  \eta_\lambda$}, where $a_{\lambda}$ and $b_{\lambda}$ are the
wavelength dependent parameters given in \cite{Cardelli1989} and
\cite{ODonnell94}, and a single \Rv\ is fit over all bands. This fit
applied to $\delta A_{\lambda}^{Si+Ca}$ leads to an average
$\Rv=2.78\pm0.34$ for the SN host galaxies in our sample. This value
is compatible with the Milky Way average of $\Rv=3.1$ The quoted
uncertainty is statistical and derived with a jackknife procedure,
removing one supernova at a time.

We tested the robustness of our \Rv\ determination in several
ways. Since the $\vec{s_\lambda}$ are measured in sequence, each one
using the corrected magnitudes from the previous step, we swap the
order in which the correction is applied, and find $\Rv=2.70$. Using
the whole sample to compute the $\vec{s_\lambda}$ instead of applying
a $2\sigma$ clipping cut leads to $\Rv=2.79$, showing the small
influence of clipping at this stage.  Host galaxy subtraction is
performed using the full spatio-spectral information from the host
obtained after the \snia\ has faded, and we do not see evidence for
residual host galaxy features in our spectra.  The measurement
covariance matrix $C_i$ is dependent on the assumed calibration
accuracy: for the present result, this is estimated from repeated
measurements of standard stars.  Another estimate can be obtained from
the residuals to a SALT2 light-curve fit. Using this in $C_i$ values
leads to $\Rv=2.76$.  The stability of the result with respect to our
choice of bandpasses is checked by removing one band at a time, and we
obtain a RMS of $\pm0.23$ around the mean result.  Finally, a Monte
Carlo simulation was performed using the converged $\gamma_\lambda$,
$\delta A_{V,i}^*$, $\eta_\lambda$ and dispersion matrix $D$ as an
input, and setting $\Rv=2.78$. The $\delta A_{\lambda,i}$ are then
randomly generated adding Gaussian noise with covariance $C_i+D$. Over
$100$ generations, the D matrix is recovered with a maximal bias of
$5\%$ on the diagonal elements and the mean fitted value is
$\Rv=2.68\pm0.03$, indicating that, if anything, our method slightly
underestimates \Rv.
  

\section{Discussion}
\label{discussion}

In this work, we selected \EWSi\ as a first variable for correction as
it provides a good proxy for $x_1$, and it is a model-independent
variable. In our data-set the Pearson correlation coefficient of
\EWSi\ with $x_{1}$ is $-0.83\pm0.04$ (Fig.~\ref{EWSi4000_vs_mag}),
confirming the result found in \cite{Arsenijevic2008}. Computing the
Hubble residuals in the B band using $(x_{1},c)$ or $(\EWSi,c)$ to
standardize SNe both lead to a dispersion of residuals of $0.16$ mag,
also confirming this result. We find that \EWSi\ is uncorrelated with
$c$ ($\rho=-0.04\pm0.12$, Fig.~\ref{EWSi4000_vs_mag}), as opposed to
the recent claim by \cite{Nordin10a}. As \EWSi\ and $x_1$ are highly
correlated, it is possible to redo the analysis with $x_{1}$ in place
of \EWSi.  The resulting SED correction vector $\vec{s_\lambda^{x_1}}$
is similar to $\vec{s_\lambda^\mathrm{Si}}$, and the conclusions
identical, yielding $\Rv=2.69$.

Computing the Hubble residuals with $(\EWSi,\EWCa,c)$ to standardize
supernovae leads to a dispersion of $0.15$~mag, which is a small
improvement. Indeed, we do observe a correlation of \EWCa\ and $c$,
with $\rho=0.34\pm0.10$.  This correlation is even increased to
$\rho=0.50$ when $c$ is computed with a U band in addition to BVR in
the light-curve fit. This shows that $c$ contains an intrinsic
component, and that the SALT2 analysis is already accounting for some
of this Ca effect. This is illustrated in
Fig.~\ref{cardellifit}\textcolor{blue}{b}: the SALT2 color model is
derived taking into account a variability measured by one intrinsic
parameter only, and thus can be compared to
$\gamma_\lambda^\mathrm{Si}$.  The observed U band contribution in
broad filters explains the UV rise in the empirical color models.  The
presence of a second intrinsic parameter induces an increased
variability in the UV as was already noticed by \cite{Ellis2008}, but
without a clear attribution to the \CaHK\ line.

Much lower effective \Rv\ values have been found previously: $\Rv=1.1$
in \cite{Tripp98}, $\Rv=2.2$ in \cite{Kessler09,Guy10} and
$\Rv\approx1-2$ in \cite{Folatelli10}. These values were derived
accounting only for one intrinsic parameter beyond color.  If we
derive an effective value after the sole \EWSi\ correction to mimic
these analyses, we obtain $\Rv=3.1$, so the explanation for the
difference has little to do with the number of intrinsic parameters
entering the correction.  This difference is explained by the
assumption on the dispersion matrix. Indeed, if we instead use the one
of \cite{Guy10}, corresponding to an RMS between $0.09$ and $0.11$~mag
on the diagonal, and all off-diagonal terms with an identical RMS of
$0.09$~mag, we obtain $\Rv=1.86$. This value is representative of
previous analyses, but the matrix used is incompatible with our best
fit matrix.  $D$ has diagonal values of 0.07, 0.04, 0.05, 0.05 and
0.10 for UBVRI respectively, lower than the values quoted by
\cite{Guy10}. However, while adjacent bands are correlated in our
matrix, an anti-correlation growing to $-1$ when the wavelength
difference increases is observed for other bands, which implies a
large color dispersion. As shown by \cite{Freedman09} and
\cite{Kessler09}, increasing the color dispersion leads to higher \Rv\
value. This long-range anti-correlation implies that uncorrected
variability in the SN spectra and/or the reddening law remains.


\section{Conclusions}
\label{conclusion}

Equivalent widths are essentially independent of host reddening and
provide a handle on the intrinsic properties of the \sneia. In
particular, \EWSi\ measured at maximum light is confirmed to be as
powerful as stretch.  After correction for this intrinsic contribution
to the magnitude, we have derived a reddening law, and proposed a
natural method to assess the dispersion due to additional
fluctuations. We found that this empirical reddening law is affected
by features linked to the SN physics, and showed that the \CaHK\ line
provides a second intrinsic variable uncorrelated with \EWSi\ or
$x_{1}$.  Correcting Hubble residuals with \EWSi\ and \EWCa\ leads to
a reddening law consistent with a canonical CCM extinction law, while
adding a dispersion in colors during the fit leads to a value of \Rv\
close to the Milky~Way value of $3.1$. Due to the coupling of \Rv\
with the residual dispersion matrix, the derived value of \Rv\ may be
affected as the remaining variability becomes better understood and
corrected.  Our plan to study the same \sneia\ over a range of phases
should help address this issue. The evolution of \sneia\ will be one
of the main limitations to their use in precision cosmology surveys at
high redshift. Our findings show that the accurate measurements of the
\CaHK\ line provides an additional tool to improve the separation of
dust and intrinsic variability. The presence of remaining scatter
offers the possibility of improvements in the future.

\begin{acknowledgements} 
  We are grateful to the technical and scientific staff of the
  University of Hawaii 2.2-meter telescope, Palomar Observatories, and
  the High Performance Research and Education Network (HPWREN) for
  their assistance in obtaining these data. We also thank the people
  of Hawaii for access to Mauna Kea. This work was based in part on
  observations with UCO facilities (Keck and Lick 3m) and NOAO
  facilities (Gemini-S (program GS-2008B-Q-26), SOAR, and CTIO 4m). We
  thank UCO and NOAO for their generous allocations of telescope
  time. We thank Julien Guy for fruitful discussions on color law
  derivation as well as the anonymous referee and Alex Kim for
  constructive comments on the text. This work was supported in France
  by CNRS/IN2P3, CNRS/INSU, CNRS/PNC, and used the resources of the
  IN2P3 computer center; This work was supported by the DFG through
  TRR33 "The Dark Universe", and by National Natural Science
  Foundation of China (grant 10903010). This work was also supported
  in part by the Director, Office of Science, Office of High Energy
  and Nuclear Physics and the Office of Advanced Scientific Computing
  Research, of the U.S.  Department of Energy (DOE) under Contract
  Nos.  DE-FG02-92ER40704, DE-AC02-05CH11231, DE-FG02-06ER06-04, and
  DE-AC02-05CH11231; by a grant from the Gordon \& Betty Moore
  Foundation; by National Science Foundation Grant Nos.  AST-0407297
  (QUEST), and 0087344 \& 0426879 (HPWREN); by a Henri Chretien
  International Research Grant administrated by the American
  Astronomical Society; the France-Berkeley Fund; by an Explora Doc
  Grant by the Region Rhone Alpes; and the Aspen Center for Physics.
\end{acknowledgements}

\bibliographystyle{aa}
\bibliography{letter-intrinsic_extrinsic-ewsi}

\begin{thebibliography}{19}
\expandafter\ifx\csname natexlab\endcsname\relax\def\natexlab#1{#1}\fi

\bibitem[{{Aldering} {et~al.}(2002){Aldering}, {Adam}, {Antilogus}, {Astier},
  {Bacon}, {Bongard}, {Bonnaud}, {Copin}, {Hardin}, {Henault}, {Howell},
  {Lemonnier}, {Levy}, {Loken}, {Nugent}, {Pain}, {Pecontal}, {Pecontal},
  {Perlmutter}, {Quimby}, {Schahmaneche}, {Smadja}, \&
  {Wood-Vasey}}]{Aldering2002}
{Aldering}, G., {Adam}, G., {Antilogus}, P., {et~al.} 2002, SPIE Conference,
  4836, 61

\bibitem[{{Arsenijevic} {et~al.}(2008){Arsenijevic}, {Fabbro}, {Mour{\~a}o},
  {Rica da Silva}, {Noname}, \& {Noname}}]{Arsenijevic2008}
{Arsenijevic}, V., {Fabbro}, S., {Mour{\~a}o}, A.~M., {et~al.} 2008, \aap, 492,
  535

\bibitem[{{Bailey} {et~al.}(2009){Bailey}, {Aldering}, {Antilogus}, {Aragon},
  {Baltay}, {Bongard}, {Buton}, {Childress}, {Chotard}, {Copin}, {Gangler},
  {Loken}, {Nugent}, {Pain}, {Pecontal}, {Pereira}, {Perlmutter}, {Rabinowitz},
  {Rigaudier}, {Runge}, {Scalzo}, {Smadja}, {Swift}, {Tao}, {Thomas}, {Wu}, \&
  {The Nearby Supernova Factory}}]{bailey2009}
{Bailey}, S., {Aldering}, G., {Antilogus}, P., {et~al.} 2009, \aap, 500, L17

\bibitem[{{Bronder} {et~al.}(2008){Bronder}, {Hook}, {Astier}, {Balam},
  {Balland}, {Basa}, {Carlberg}, {Conley}, {Fouchez}, {Guy}, {Howell}, {Neill},
  {Pain}, {Perrett}, {Pritchet}, {Regnault}, {Sullivan}, {Baumont}, {Fabbro},
  {Filliol}, {Perlmutter}, \& {Ripoche}}]{Bronder2008}
{Bronder}, T.~J., {Hook}, I.~M., {Astier}, P., {et~al.} 2008, \aap, 477, 717

\bibitem[{{Cardelli} {et~al.}(1989){Cardelli}, {Clayton}, \&
  {Mathis}}]{Cardelli1989}
{Cardelli}, J.~A., {Clayton}, G.~C., \& {Mathis}, J.~S. 1989, \apj, 345, 245

\bibitem[{{Ellis} {et~al.}(2008){Ellis}, {Sullivan}, {Nugent}, {Howell},
  {Gal-Yam}, {Astier}, {Balam}, {Balland}, {Basa}, {Carlberg}, {Conley},
  {Fouchez}, {Guy}, {Hardin}, {Hook}, {Pain}, {Perrett}, {Pritchet}, \&
  {Regnault}}]{Ellis2008}
{Ellis}, R.~S., {Sullivan}, M., {Nugent}, P.~E., {et~al.} 2008, \apj, 674, 51

\bibitem[{{Folatelli} {et~al.}(2010){Folatelli}, {Phillips}, {Burns},
  {Contreras}, {Hamuy}, {Freedman}, {Persson}, {Stritzinger}, {Suntzeff},
  {Krisciunas}, {Boldt}, {Gonz{\'a}lez}, {Krzeminski}, {Morrell}, {Roth},
  {Salgado}, {Madore}, {Murphy}, {Wyatt}, {Li}, {Filippenko}, \&
  {Miller}}]{Folatelli10}
{Folatelli}, G., {Phillips}, M.~M., {Burns}, C.~R., {et~al.} 2010, \aj, 139,
  120

\bibitem[{{Freedman} {et~al.}(2009){Freedman}, {Burns}, {Phillips}, {Wyatt},
  {Persson}, {Madore}, {Contreras}, {Folatelli}, {Gonzalez}, {Hamuy}, {Hsiao},
  {Kelson}, {Morrell}, {Murphy}, {Roth}, {Stritzinger}, {Sturch}, {Suntzeff},
  {Astier}, {Balland}, {Bassett}, {Boldt}, {Carlberg}, {Conley}, {Frieman},
  {Garnavich}, {Guy}, {Hardin}, {Howell}, {Kessler}, {Lampeitl}, {Marriner},
  {Pain}, {Perrett}, {Regnault}, {Riess}, {Sako}, {Schneider}, {Sullivan}, \&
  {Wood-Vasey}}]{Freedman09}
{Freedman}, W.~L., {Burns}, C.~R., {Phillips}, M.~M., {et~al.} 2009, \apj, 704,
  1036

\bibitem[{{Guy} {et~al.}(2007){Guy}, {Astier}, {Baumont}, {Hardin}, {Pain},
  {Regnault}, {Basa}, {Carlberg}, {Conley}, {Fabbro}, {Fouchez}, {Hook},
  {Howell}, {Perrett}, {Pritchet}, {Rich}, {Sullivan}, {Antilogus}, {Aubourg},
  {Bazin}, {Bronder}, {Filiol}, {Palanque-Delabrouille}, {Ripoche}, \&
  {Ruhlmann-Kleider}}]{Guy2007}
{Guy}, J., {Astier}, P., {Baumont}, S., {et~al.} 2007, \aap, 466, 11

\bibitem[{{Guy} {et~al.}(2010){Guy}, {Sullivan}, {Conley}, {Regnault},
  {Astier}, {Balland}, {Basa}, {Carlberg}, {Fouchez}, {Hardin}, {Hook},
  {Howell}, {Pain}, {Palanque-Delabrouille}, {Perrett}, {Pritchet}, {Rich},
  {Ruhlmann-Kleider}, {Balam}, {Baumont}, {Ellis}, {Fabbro}, {Fakhouri},
  {Fourmanoit}, {Gonz{\'a}lez-Gait{\'a}n}, {Graham}, {Hsiao}, {Kronborg},
  {Lidman}, {Mourao}, {Perlmutter}, {Ripoche}, {Suzuki}, \& {Walker}}]{Guy10}
{Guy}, J., {Sullivan}, M., {Conley}, A., {et~al.} 2010, \aap, 523, A7+

\bibitem[{{Hicken} {et~al.}(2009){Hicken}, {Challis}, {Jha}, {Kirshner},
  {Matheson}, {Modjaz}, {Rest}, {Michael Wood-Vasey}, {Bakos}, {Barton},
  {Berlind}, {Bragg}, {Brice{\~n}o}, {Brown}, {Caldwell}, {Calkins}, {Cho},
  {Ciupik}, {Contreras}, {Dendy}, {Dosaj}, {Durham}, {Eriksen}, {Esquerdo},
  {Everett}, {Falco}, {Fernandez}, {Gaba}, {Garnavich}, {Graves}, {Green},
  {Groner}, {Hergenrother}, {Holman}, {Hradecky}, {Huchra}, {Hutchison},
  {Jerius}, {Jordan}, {Kilgard}, {Krauss}, {Luhman}, {Macri}, {Marrone},
  {McDowell}, {McIntosh}, {McNamara}, {Megeath}, {Mochejska}, {Munoz},
  {Muzerolle}, {Naranjo}, {Narayan}, {Pahre}, {Peters}, {Peterson}, {Rines},
  {Ripman}, {Roussanova}, {Schild}, {Sicilia-Aguilar}, {Sokoloski}, {Smalley},
  {Smith}, {Spahr}, {Stanek}, {Barmby}, {Blondin}, {Stubbs}, {Szentgyorgyi},
  {Torres}, {Vaz}, {Vikhlinin}, {Wang}, {Westover}, {Woods}, \&
  {Zhao}}]{Hicken09a}
{Hicken}, M., {Challis}, P., {Jha}, S., {et~al.} 2009, \apj, 700, 331

\bibitem[{{Jha} {et~al.}(2007){Jha}, {Riess}, \& {Kirshner}}]{Jha2007}
{Jha}, S., {Riess}, A.~G., \& {Kirshner}, R.~P. 2007, \apj, 659, 122

\bibitem[{{Kessler} {et~al.}(2009){Kessler}, {Becker}, {Cinabro}, {Vanderplas},
  {Frieman}, {Marriner}, {Davis}, {Dilday}, {Holtzman}, {Jha}, {Lampeitl},
  {Sako}, {Smith}, {Zheng}, {Nichol}, {Bassett}, {Bender}, {Depoy}, {Doi},
  {Elson}, {Filippenko}, {Foley}, {Garnavich}, {Hopp}, {Ihara}, {Ketzeback},
  {Kollatschny}, {Konishi}, {Marshall}, {McMillan}, {Miknaitis}, {Morokuma},
  {M{\"o}rtsell}, {Pan}, {Prieto}, {Richmond}, {Riess}, {Romani}, {Schneider},
  {Sollerman}, {Takanashi}, {Tokita}, {van der Heyden}, {Wheeler}, {Yasuda}, \&
  {York}}]{Kessler09}
{Kessler}, R., {Becker}, A.~C., {Cinabro}, D., {et~al.} 2009, \apjs, 185, 32

\bibitem[{{Nordin} {et~al.}(2010){Nordin}, {Ostman}, {Goobar}, {Balland},
  {Lampeitl}, {Nichol}, {Sako}, {Schneider}, {Smith}, {Sollerman}, \&
  {Wheeler}}]{Nordin10a}
{Nordin}, J., {Ostman}, L., {Goobar}, A., {et~al.} 2010, ArXiv
  e-prints:1012.4430

\bibitem[{{O'Donnell}(1994)}]{ODonnell94}
{O'Donnell}, J.~E. 1994, \apj, 422, 158

\bibitem[{{Phillips}(1993)}]{Phillips1993}
{Phillips}, M.~M. 1993, \apjl, 413, L105

\bibitem[{{Tripp}(1998)}]{Tripp98}
{Tripp}, R. 1998, \aap, 331, 815

\bibitem[{{Walker} {et~al.}(2010){Walker}, {Hook}, {Sullivan}, {Howell},
  {Astier}, {Balland}, {Basa}, {Bronder}, {Carlberg}, {Conley}, {Fouchez},
  {Guy}, {Hardin}, {Pain}, {Perrett}, {Pritchet}, {Regnault}, {Rich},
  {Aldering}, {Fakhouri}, {Kronborg}, {Palanque-Delabrouille}, {Perlmutter},
  {Ruhlmann-Kleider}, \& {Zhang}}]{Walker2010}
{Walker}, E.~S., {Hook}, I.~M., {Sullivan}, M., {et~al.} 2010, \mnras, 1811

\bibitem[{{Wang} {et~al.}(2009){Wang}, {Filippenko}, {Ganeshalingam}, {Li},
  {Silverman}, {Wang}, {Chornock}, {Foley}, {Gates}, {Macomber}, {Serduke},
  {Steele}, \& {Wong}}]{Wang09b}
{Wang}, X., {Filippenko}, A.~V., {Ganeshalingam}, M., {et~al.} 2009, \apjl,
  699, L139

\end{thebibliography}

\scriptsize
\onllongtabL{1}{
  \begin{landscape}
    \begin{longtable}{lrrrrrrrrrrr}
      \caption{Supernovae used in this work, including synthetic
        magnitudes after phase correction $\delta M_{\lambda}$, SALT2
        fit parameters $x_1$ and $c$, equivalent widths \EWSi\ and
        \EWCa\ in \AA\ and phases in days.
        \label{datatable}
      } \\ \hline \hline
      Name
      & \multicolumn{1}{c}{$z_{cmb}^{1}$}
      & \multicolumn{1}{c}{$\delta M_{\mathrm{U}}$}
      & \multicolumn{1}{c}{$\delta M_{\mathrm{B}}$}
      & \multicolumn{1}{c}{$\delta M_{\mathrm{V}}$}
      & \multicolumn{1}{c}{$\delta M_{\mathrm{R}}$}
      & \multicolumn{1}{c}{$\delta M_{\mathrm{I}}$}
      & \multicolumn{1}{c}{$x_1$}
      & \multicolumn{1}{c}{$c$}
      & \multicolumn{1}{c}{\EWSi}
      & \multicolumn{1}{c}{\EWCa}
      & \multicolumn{1}{c}{phase} \\
      \hline
      \endfirsthead
      \caption{continued.} \\ \hline \hline
      Name
      & \multicolumn{1}{c}{$z_{cmb}^{1}$}
      & \multicolumn{1}{c}{$\delta M_{\mathrm{U}}$}
      & \multicolumn{1}{c}{$\delta M_{\mathrm{B}}$}
      & \multicolumn{1}{c}{$\delta M_{\mathrm{V}}$}
      & \multicolumn{1}{c}{$\delta M_{\mathrm{R}}$}
      & \multicolumn{1}{c}{$\delta M_{\mathrm{I}}$}
      & \multicolumn{1}{c}{$x_1$}
      & \multicolumn{1}{c}{$c$}
      & \multicolumn{1}{c}{\EWSi}
      & \multicolumn{1}{c}{\EWCa}
      & \multicolumn{1}{c}{phase} \\
      \hline
      \endhead
      \endfoot
      
SN2004ef$^{a}$ & $0.0298$ & $0.408 \pm 0.012$ & $0.357 \pm 0.010$ & $0.249 \pm 0.010$ & $0.208 \pm 0.009$ & $0.148 \pm 0.010$ & $-1.62 \pm 0.29$ & $0.181 \pm 0.024$ & $22.7 \pm 1.5$ & $140.4 \pm 2.5$ & $0.5 \pm 0.2$ \\ 
SNF20050728-006 & $0.0582$ & $0.428 \pm 0.065$ & $0.322 \pm 0.064$ & $0.239 \pm
0.064$ & $0.202 \pm 0.063$ & $0.161 \pm 0.064$ & $0.85 \pm 0.53$ & $0.159 \pm 0.027$ & $13.9 \pm 2.5$ & $144.6 \pm 6.2$ & $-0.9 \pm 0.4$
\\ 
SNF20051022-006 & $0.0449$ & $-0.153 \pm 0.051$ & $-0.124 \pm 0.050$ & $-0.096 \pm 0.051$ & $-0.013 \pm 0.050$ & $0.030 \pm 0.051$ & $0.95 \pm 0.35$ & $0.000 \pm 0.021$ & $9.6 \pm 1.1$ & $127.5 \pm 2.4$ & $0.7 \pm 0.4$ \\ 
SNF20060511-014 & $0.0466$ & $-0.002 \pm 0.085$ & $-0.079 \pm 0.085$ & $-0.029 \pm 0.085$ & $-0.102 \pm 0.084$ & $-0.022 \pm 0.086$ & $-1.27 \pm 0.15$ & $0.022 \pm 0.022$ & $16.5 \pm 1.4$ & $101.3 \pm 11.2$ & $0.6 \pm 0.1$ \\ 
SNF20060526-003 & $0.0787$ & $-0.068 \pm 0.025$ & $-0.049 \pm 0.023$ & $-0.066 \pm 0.023$ & $-0.009 \pm 0.022$ & $-0.004 \pm 0.023$ & $-0.90 \pm 0.27$ & $0.013 \pm 0.017$ & $11.4 \pm 1.9$ & $121.3 \pm 6.1$ & $-0.8 \pm 0.3$ \\ 
SNF20060621-015 & $0.0541$ & $-0.443 \pm 0.027$ & $-0.368 \pm 0.026$ & $-0.300
\pm 0.026$ & $-0.218 \pm 0.026$ & $-0.189 \pm 0.026$ & $0.09 \pm 0.21$
& $-0.071 \pm 0.015$ & $11.0 \pm 1.5$ & $106.9 \pm 7.7$ & $1.1 \pm
0.3$ \\ 
SN2006dm$^{b}$ & $0.0208$ & $0.381 \pm 0.017$ & $0.368 \pm 0.015$ & $0.331 \pm
0.015$ & $0.295 \pm 0.014$ & $0.275 \pm 0.015$ & $-2.13 \pm 0.34$ &
$0.084 \pm 0.030$ & $31.1 \pm 1.2$ & $101.1 \pm 2.7$ & $-1.3 \pm 0.2$
\\ 
SN2006do$^{b}$ & $0.0274$ & $0.163 \pm 0.014$ & $0.170 \pm 0.012$ & $0.152 \pm 0.012$ & $0.149 \pm 0.012$ & $0.253 \pm 0.034$ & $-2.86 \pm 0.68$ & $0.003 \pm 0.036$ & $24.6 \pm 1.0$ & $117.7 \pm 2.5$ & $-0.6 \pm 0.4$ \\ 
SNF20060911-014 & $0.0887$ & $-0.468 \pm 0.126$ & $-0.365 \pm 0.126$ & $-0.318 \pm 0.126$ & $-0.228 \pm 0.126$ & $-0.187 \pm 0.126$ & $1.02 \pm 0.21$ & $-0.040 \pm 0.015$ & $4.3 \pm 1.4$ & $80.3 \pm 9.4$ & $-0.8 \pm 0.4$ \\ 
SNF20060919-007 & $0.0694$ & $-0.229 \pm 0.039$ & $-0.168 \pm 0.037$ & $-0.126 \pm 0.038$ & $-0.098 \pm 0.037$ & $-0.069 \pm 0.038$ & $-0.52 \pm 0.29$ & $-0.012 \pm 0.018$ & $11.5 \pm 1.2$ & $125.9 \pm 2.4$ & $-0.4 \pm 0.7$ \\ 
SNF20061020-000 & $0.0372$ & $0.252 \pm 0.069$ & $0.264 \pm 0.068$ & $0.269 \pm 0.068$ & $0.146 \pm 0.068$ & $0.044 \pm 0.068$ & $-1.75 \pm 0.27$ & $0.103 \pm 0.031$ & $25.4 \pm 1.6$ & $97.3 \pm 8.4$ & $0.9 \pm 0.5$ \\ 
SNF20061021-003 & $0.0615$ & $0.190 \pm 0.038$ & $0.114 \pm 0.037$ & $-0.017 \pm 0.037$ & $-0.016 \pm 0.037$ & $-0.078 \pm 0.038$ & $-0.13 \pm 0.14$ & $0.118 \pm 0.013$ & $9.0 \pm 1.9$ & $123.7 \pm 2.7$ & $-0.7 \pm 0.0$ \\ 
SNF20061024-000 & $0.0557$ & $0.464 \pm 0.054$ & $0.365 \pm 0.051$ & $0.376 \pm 0.051$ & $0.298 \pm 0.051$ & $0.290 \pm 0.052$ & $-2.51 \pm 0.31$ & $0.040 \pm 0.041$ & $28.6 \pm 2.3$ & $86.3 \pm 27.4$ & $0.3 \pm 0.6$ \\ 
SNF20061108-004 & $0.0879$ & $-0.265 \pm 0.020$ & $-0.274 \pm 0.017$ & $-0.303 \pm 0.017$ & $-0.240 \pm 0.017$ & $-0.072 \pm 0.039$ & $0.58 \pm 0.20$ & $0.043 \pm 0.014$ & $12.8 \pm 4.7$ & $129.1 \pm 6.7$ & $-0.2 \pm 0.0$ \\ 
SNF20061111-002 & $0.0682$ & $0.099 \pm 0.035$ & $-0.046 \pm 0.034$ & $-0.068 \pm
0.034$ & $-0.120 \pm 0.034$ & $-0.102 \pm 0.047$ & $1.05 \pm 0.31$ &
$0.132 \pm 0.018$ & $22.6 \pm 2.1$ & $112.5 \pm 8.4$ & $-0.9 \pm 0.3$
\\ 
SN2006ob$^{c}$ & $0.0582$ & $0.114 \pm 0.017$ & $0.196 \pm 0.014$ & $0.226 \pm 0.015$ & $0.174 \pm 0.014$ & $0.127 \pm 0.015$ & $-3.11 \pm 0.44$ & $0.006 \pm 0.034$ & $28.5 \pm 1.8$ & $92.0 \pm 14.0$ & $0.0 \pm 0.4$ \\ 
SNF20070403-001 & $0.0815$ & $-0.046 \pm 0.015$ & $-0.008 \pm 0.012$ & $-0.027
\pm 0.013$ & $-0.019 \pm 0.012$ & $0.079 \pm 0.036$ & $-1.56 \pm 0.20$
& $0.032 \pm 0.014$ & $20.9 \pm 2.6$ & $97.9 \pm 16.4$ & $-0.0 \pm
0.3$ \\ 
SN2007bd$^{d}$ & $0.0320$ & $-0.202 \pm 0.021$ & $-0.162 \pm 0.019$ & $-0.097 \pm 0.019$ & $-0.093 \pm 0.019$ & $-0.108 \pm 0.037$ & $-1.12 \pm 0.37$ & $-0.036 \pm 0.024$ & $19.0 \pm 1.0$ & $108.8 \pm 8.8$ & $-0.4 \pm 0.4$ \\ 
SNF20070424-003 & $0.0680$ & $0.138 \pm 0.045$ & $0.063 \pm 0.044$ & $0.102 \pm 0.044$ & $0.088 \pm 0.044$ & $0.024 \pm 0.055$ & $0.09 \pm 0.11$ & $0.029 \pm 0.013$ & $14.2 \pm 2.1$ & $128.5 \pm 4.9$ & $0.5 \pm 0.1$ \\ 
SNF20070427-001 & $0.0778$ & $-0.761 \pm 0.016$ & $-0.521 \pm 0.013$ & $-0.352 \pm 0.014$ & $-0.235 \pm 0.013$ & $-0.228 \pm 0.015$ & $-0.73 \pm 0.13$ & $-0.173 \pm 0.012$ & $7.7 \pm 1.3$ & $84.8 \pm 3.1$ & $-0.5 \pm 0.3$ \\ 
SNF20070506-006 & $0.0355$ & $-0.471 \pm 0.063$ & $-0.404 \pm 0.063$ & $-0.407 \pm 0.063$ & $-0.329 \pm 0.063$ & $-0.308 \pm 0.063$ & $1.00 \pm 0.14$ & $0.021 \pm 0.017$ & $7.7 \pm 0.8$ & $95.3 \pm 1.9$ & $-0.1 \pm 0.1$ \\ 
SNF20070531-011 & $0.0358$ & $0.078 \pm 0.060$ & $0.047 \pm 0.059$ & $0.085 \pm
0.059$ & $0.072 \pm 0.059$ & $0.137 \pm 0.063$ & $-2.71 \pm 0.43$ &
$-0.067 \pm 0.040$ & $23.9 \pm 1.3$ & $112.4 \pm 13.8$ & $0.3 \pm 0.6$
\\ 
SN2007cq$^{e}$ & $0.0246$ & $-0.299 \pm 0.082$ & $-0.260 \pm 0.082$ & $-0.279 \pm 0.082$ & $-0.198 \pm 0.082$ & $-0.228 \pm 0.082$ & $-0.82 \pm 0.18$ & $-0.013 \pm 0.019$ & $12.8 \pm 2.1$ & $66.7 \pm 6.4$ & $-2.3 \pm 0.2$ \\ 
SNF20070630-006 & $0.0706$ & $-0.072 \pm 0.029$ & $-0.080 \pm 0.028$ & $-0.021 \pm 0.028$ & $-0.035 \pm 0.028$ & $-0.029 \pm 0.028$ & $0.35 \pm 0.13$ & $0.027 \pm 0.013$ & $10.3 \pm 3.6$ & $139.4 \pm 5.1$ & $1.1 \pm 0.2$ \\ 
SNF20070701-005 & $0.0683$ & $-0.311 \pm 0.045$ & $-0.261 \pm 0.044$ & $-0.250 \pm 0.044$ & $-0.263 \pm 0.044$ & $-0.178 \pm 0.044$ & $-0.89 \pm 0.24$ & $0.021 \pm 0.019$ & $12.9 \pm 1.4$ & $105.0 \pm 2.8$ & $0.0 \pm 0.0$ \\ 
SNF20070712-000 & $0.0296$ & $0.166 \pm 0.047$ & $0.111 \pm 0.046$ & $0.079 \pm 0.046$ & $0.080 \pm 0.045$ & $0.140 \pm 0.046$ & $-0.78 \pm 0.17$ & $0.060 \pm 0.022$ & $19.5 \pm 1.8$ & $93.1 \pm 8.2$ & $2.2 \pm 0.0$ \\ 
SNF20070712-003 & $0.0738$ & $-0.240 \pm 0.042$ & $-0.221 \pm 0.041$ & $-0.130 \pm 0.041$ & $-0.104 \pm 0.041$ & $-0.148 \pm 0.041$ & $0.08 \pm 0.17$ & $-0.025 \pm 0.013$ & $14.4 \pm 1.8$ & $104.2 \pm 12.3$ & $-1.8 \pm 0.2$ \\ 
SNF20070717-003 & $0.0859$ & $0.210 \pm 0.022$ & $0.122 \pm 0.020$ & $0.175 \pm 0.020$ & $0.105 \pm 0.019$ & $0.107 \pm 0.021$ & $-1.42 \pm 0.22$ & $0.060 \pm 0.016$ & $24.3 \pm 3.0$ & $112.1 \pm 41.9$ & $1.9 \pm 0.2$ \\ 
SNF20070725-001 & $0.0668$ & $-0.413 \pm 0.043$ & $-0.353 \pm 0.042$ & $-0.253 \pm 0.042$ & $-0.152 \pm 0.042$ & $-0.056 \pm 0.043$ & $0.40 \pm 0.15$ & $-0.086 \pm 0.015$ & $11.9 \pm 1.5$ & $107.2 \pm 2.6$ & $0.6 \pm 0.4$ \\ 
SNF20070727-016 & $0.0665$ & $-0.817 \pm 0.029$ & $-0.593 \pm 0.027$ & $-0.517 \pm 0.027$ & $-0.437 \pm 0.027$ & $-0.390 \pm 0.028$ & $0.08 \pm 0.15$ & $-0.053 \pm 0.015$ & $4.9 \pm 1.6$ & $80.6 \pm 3.3$ & $1.0 \pm 0.5$ \\ 
SNF20070802-000 & $0.0643$ & $0.360 \pm 0.041$ & $0.208 \pm 0.041$ & $0.144 \pm 0.041$ & $0.128 \pm 0.040$ & $0.150 \pm 0.041$ & $-0.07 \pm 0.16$ & $0.124 \pm 0.012$ & $19.2 \pm 2.3$ & $159.7 \pm 4.9$ & $0.7 \pm 0.2$ \\ 
SNF20070803-005 & $0.0303$ & $-0.647 \pm 0.146$ & $-0.459 \pm 0.146$ & $-0.487 \pm 0.146$ & $-0.422 \pm 0.146$ & $-0.455 \pm 0.146$ & $0.53 \pm 0.19$ & $0.062 \pm 0.019$ & $1.3 \pm 2.3$ & $38.8 \pm 4.1$ & $-2.0 \pm 0.1$ \\ 
SNF20070806-026 & $0.0440$ & $-0.010 \pm 0.064$ & $0.052 \pm 0.063$ & $0.118 \pm 0.063$ & $0.069 \pm 0.063$ & $0.076 \pm 0.063$ & $-2.18 \pm 0.18$ & $-0.080 \pm 0.033$ & $27.1 \pm 1.9$ & $98.4 \pm 11.1$ & $1.8 \pm 0.2$ \\ 
SNF20070810-004 & $0.0827$ & $-0.015 \pm 0.030$ & $-0.097 \pm 0.029$ & $-0.060
\pm 0.029$ & $-0.026 \pm 0.029$ & $0.031 \pm 0.044$ & $-0.53 \pm 0.11$
& $0.001 \pm 0.011$ & $22.9 \pm 2.3$ & $127.8 \pm 4.8$ & $2.2 \pm 0.0$
\\ 
SNF20070818-001 & $0.0728$ & $0.269 \pm 0.041$ & $0.059 \pm 0.039$ & $-0.011 \pm 0.039$ & $-0.017 \pm 0.038$ & $0.046 \pm 0.039$ & $-0.51 \pm 0.16$ & $0.101 \pm 0.013$ & $17.8 \pm 2.9$ & $156.0 \pm 9.1$ & $-1.2 \pm 0.2$ \\ 
SNF20070817-003 & $0.0630$ & $0.227 \pm 0.176$ & $0.129 \pm 0.176$ & $0.190 \pm
0.176$ & $0.184 \pm 0.176$ & $0.273 \pm 0.179$ & $-0.95 \pm 0.15$ &
$-0.020 \pm 0.014$ & $18.6 \pm 2.1$ & $98.0 \pm 4.3$ & $1.7 \pm 0.2$
\\ 
SN2007kk$^{f}$ & $0.0406$ & $-0.231 \pm 0.017$ & $-0.271 \pm 0.015$ & $-0.278 \pm 0.015$ & $-0.208 \pm 0.014$ & $-0.128 \pm 0.015$ & $0.06 \pm 0.19$ & $0.042 \pm 0.015$ & $10.5 \pm 1.2$ & $128.6 \pm 2.1$ & $-0.4 \pm 0.3$ \\ 
SNF20071015-000 & $0.0373$ & $1.340 \pm 0.065$ & $1.166 \pm 0.064$ & $0.876 \pm
0.064$ & $0.672 \pm 0.064$ & $0.410 \pm 0.064$ & $0.79 \pm 0.35$ &
$0.481 \pm 0.025$ & $6.5 \pm 1.4$ & $102.8 \pm 6.5$ & $-1.0 \pm 0.4$
\\ 
SN2007le$^{g}$ & $0.0055$ & $1.445 \pm 0.016$ & $1.109 \pm 0.014$ & $0.817 \pm 0.014$ & $0.723 \pm 0.013$ & $0.675 \pm 0.014$ & $0.36 \pm 0.26$ & $0.417 \pm 0.027$ & $14.3 \pm 1.0$ & $149.0 \pm 2.8$ & $-1.2 \pm 0.2$ \\ 
SNF20071021-000 & $0.0262$ & $0.471 \pm 0.034$ & $0.238 \pm 0.034$ & $0.166 \pm
0.034$ & $0.161 \pm 0.034$ & $0.220 \pm 0.034$ & $-0.88 \pm 0.21$ &
$0.120 \pm 0.020$ & $20.5 \pm 1.1$ & $165.2 \pm 8.0$ & $-0.6 \pm 0.1$
\\ 
SN2007nq$^{h}$ & $0.0439$ & $0.038 \pm 0.022$ & $0.094 \pm 0.019$ & $0.084 \pm 0.019$ & $0.038 \pm 0.018$ & $0.015 \pm 0.020$ & $-2.56 \pm 0.43$ & $0.068 \pm 0.043$ & $24.5 \pm 1.7$ & $89.7 \pm 8.5$ & $1.5 \pm 0.7$ \\ 
SNF20080323-009 & $0.0834$ & $-0.383 \pm 0.017$ & $-0.201 \pm 0.015$ & $-0.081 \pm 0.015$ & $-0.082 \pm 0.014$ & $-0.118 \pm 0.017$ & $-0.32 \pm 0.18$ & $-0.033 \pm 0.016$ & $13.4 \pm 1.6$ & $105.0 \pm 4.0$ & $0.6 \pm 0.4$ \\ 
SNF20080510-001 & $0.0723$ & $-0.166 \pm 0.028$ & $-0.172 \pm 0.027$ & $-0.116 \pm 0.028$ & $-0.075 \pm 0.027$ & $-0.039 \pm 0.028$ & $-0.29 \pm 0.15$ & $-0.028 \pm 0.012$ & $18.5 \pm 2.9$ & $117.7 \pm 2.8$ & $-0.7 \pm 0.1$ \\ 
SNF20080512-010 & $0.0639$ & $-0.149 \pm 0.196$ & $-0.023 \pm 0.196$ & $0.044 \pm 0.196$ & $0.027 \pm 0.196$ & $0.021 \pm 0.196$ & $-2.21 \pm 0.28$ & $-0.026 \pm 0.026$ & $23.9 \pm 1.5$ & $94.8 \pm 4.7$ & $0.3 \pm 0.1$ \\ 
SNF20080514-002 & $0.0230$ & $-0.130 \pm 0.012$ & $0.018 \pm 0.011$ & $0.098 \pm 0.011$ & $0.117 \pm 0.011$ & $0.078 \pm 0.026$ & $-1.99 \pm 0.15$ & $-0.075 \pm 0.020$ & $18.8 \pm 0.5$ & $81.0 \pm 1.7$ & $-0.1 \pm 0.1$ \\ 
SNF20080516-000 & $0.0736$ & $-0.395 \pm 0.038$ & $-0.306 \pm 0.038$ & $-0.290 \pm 0.038$ & $-0.204 \pm 0.037$ & $-0.214 \pm 0.038$ & $1.43 \pm 0.24$ & $0.008 \pm 0.018$ & $4.6 \pm 1.5$ & $104.9 \pm 3.5$ & $-1.2 \pm 0.4$ \\ 
SNF20080516-022 & $0.0747$ & $-0.518 \pm 0.016$ & $-0.345 \pm 0.014$ & $-0.184 \pm 0.014$ & $-0.161 \pm 0.013$ & $-0.246 \pm 0.016$ & $-0.34 \pm 0.11$ & $-0.096 \pm 0.012$ & $13.8 \pm 2.4$ & $89.2 \pm 15.4$ & $0.2 \pm 0.1$ \\ 
SNF20080522-000 & $0.0463$ & $-0.622 \pm 0.060$ & $-0.423 \pm 0.059$ & $-0.480 \pm 0.059$ & $-0.411 \pm 0.059$ & $-0.467 \pm 0.059$ & $1.09 \pm 0.17$ & $0.093 \pm 0.017$ & $4.3 \pm 2.2$ & $46.5 \pm 25.0$ & $0.7 \pm 0.2$ \\ 
SNF20080522-011 & $0.0385$ & $-0.421 \pm 0.038$ & $-0.424 \pm 0.037$ & $-0.393 \pm 0.037$ & $-0.345 \pm 0.037$ & $-0.361 \pm 0.037$ & $1.28 \pm 0.24$ & $0.026 \pm 0.019$ & $8.9 \pm 0.7$ & $120.3 \pm 2.3$ & $0.6 \pm 0.3$ \\ 
SNF20080531-000 & $0.0368$ & $0.072 \pm 0.055$ & $0.021 \pm 0.055$ & $0.038 \pm 0.055$ & $0.041 \pm 0.055$ & $0.079 \pm 0.055$ & $-0.81 \pm 0.21$ & $0.014 \pm 0.017$ & $20.3 \pm 2.1$ & $132.8 \pm 2.6$ & $0.5 \pm 0.2$ \\ 
SNF20080610-000 & $0.0789$ & $0.106 \pm 0.017$ & $0.041 \pm 0.015$ & $0.106 \pm 0.015$ & $0.104 \pm 0.014$ & $0.125 \pm 0.016$ & $0.03 \pm 0.13$ & $0.015 \pm 0.013$ & $17.9 \pm 2.0$ & $119.0 \pm 4.9$ & $1.0 \pm 0.0$ \\ 
SNF20080612-003 & $0.0330$ & $-0.522 \pm 0.046$ & $-0.487 \pm 0.045$ & $-0.487 \pm 0.045$ & $-0.442 \pm 0.045$ & $-0.442 \pm 0.054$ & $0.52 \pm 0.20$ & $0.010 \pm 0.018$ & $5.9 \pm 1.0$ & $121.4 \pm 2.3$ & $-0.9 \pm 0.4$ \\ 
SNF20080614-010 & $0.0752$ & $0.109 \pm 0.018$ & $0.162 \pm 0.016$ & $0.262 \pm 0.016$ & $0.167 \pm 0.015$ & $0.133 \pm 0.016$ & $-2.90 \pm 0.24$ & $-0.006 \pm 0.026$ & $25.7 \pm 2.6$ & $99.1 \pm 36.8$ & $1.3 \pm 0.3$ \\ 
SNF20080623-001 & $0.0448$ & $0.081 \pm 0.026$ & $0.027 \pm 0.025$ & $0.094 \pm 0.025$ & $0.111 \pm 0.024$ & $0.173 \pm 0.025$ & $-0.44 \pm 0.13$ & $-0.031 \pm 0.013$ & $17.1 \pm 1.1$ & $149.5 \pm 1.8$ & $0.4 \pm 0.1$ \\ 
SNF20080626-002 & $0.0232$ & $-0.290 \pm 0.090$ & $-0.258 \pm 0.090$ & $-0.267 \pm 0.090$ & $-0.168 \pm 0.090$ & $-0.112 \pm 0.090$ & $0.83 \pm 0.24$ & $0.000 \pm 0.020$ & $5.5 \pm 3.1$ & $129.1 \pm 1.0$ & $-0.1 \pm 0.3$ \\ 
SNF20080714-008 & $0.0846$ & $0.547 \pm 0.038$ & $0.350 \pm 0.036$ & $0.253 \pm
0.036$ & $0.217 \pm 0.036$ & $0.212 \pm 0.044$ & $-0.39 \pm 0.15$ &
$0.145 \pm 0.012$ & $23.8 \pm 3.7$ & $120.4 \pm 17.4$ & $0.2 \pm 0.2$
\\ 
SN2008ec$^{d}$ & $0.0151$ & $0.503 \pm 0.011$ & $0.472 \pm 0.009$ & $0.368 \pm 0.009$ & $0.267 \pm 0.008$ & $0.127 \pm 0.009$ & $-1.64 \pm 0.18$ & $0.202 \pm 0.024$ & $23.6 \pm 0.7$ & $102.6 \pm 18.2$ & $0.6 \pm 0.1$ \\ 
SNF20080720-001 & $0.0194$ & $1.647 \pm 0.227$ & $1.184 \pm 0.227$ & $0.792 \pm 0.227$ & $0.538 \pm 0.227$ & $0.308 \pm 0.227$ & $0.61 \pm 0.17$ & $0.627 \pm 0.016$ & $12.8 \pm 3.6$ & $144.9 \pm 3.8$ & $0.6 \pm 0.3$ \\ 
SNF20080802-006 & $0.0713$ & $0.271 \pm 0.036$ & $0.297 \pm 0.035$ & $0.352 \pm 0.035$ & $0.303 \pm 0.035$ & $0.460 \pm 0.036$ & $-2.95 \pm 0.53$ & $-0.010 \pm 0.031$ & $23.8 \pm 3.0$ & $113.0 \pm 6.7$ & $-0.5 \pm 0.1$ \\ 
SNF20080803-000 & $0.0571$ & $0.303 \pm 0.019$ & $0.261 \pm 0.016$ & $0.183 \pm 0.017$ & $0.161 \pm 0.016$ & $0.173 \pm 0.017$ & $0.04 \pm 0.16$ & $0.143 \pm 0.016$ & $7.7 \pm 2.1$ & $117.9 \pm 3.9$ & $-1.1 \pm 0.1$ \\ 
SNF20080810-001 & $0.0423$ & $0.012 \pm 0.031$ & $0.089 \pm 0.030$ & $0.146 \pm 0.030$ & $0.081 \pm 0.030$ & $-0.003 \pm 0.030$ & $-1.52 \pm 0.16$ & $0.030 \pm 0.019$ & $23.8 \pm 1.0$ & $90.2 \pm 22.9$ & $-0.0 \pm 0.1$ \\ 
SNF20080822-005 & $0.0705$ & $-0.532 \pm 0.052$ & $-0.355 \pm 0.052$ & $-0.346 \pm 0.052$ & $-0.264 \pm 0.051$ & $-0.197 \pm 0.052$ & $0.35 \pm 0.26$ & $-0.005 \pm 0.017$ & $7.3 \pm 1.4$ & $79.9 \pm 3.5$ & $1.1 \pm 0.5$ \\ 
SNF20080825-010 & $0.0396$ & $-0.198 \pm 0.057$ & $-0.134 \pm 0.057$ & $-0.102 \pm 0.057$ & $-0.120 \pm 0.056$ & $-0.165 \pm 0.057$ & $-1.34 \pm 0.23$ & $0.037 \pm 0.021$ & $19.5 \pm 1.8$ & $102.8 \pm 14.1$ & $0.8 \pm 0.2$ \\ 
SNF20080909-030 & $0.0298$ & $-0.139 \pm 0.033$ & $-0.139 \pm 0.032$ & $-0.166 \pm 0.032$ & $-0.115 \pm 0.032$ & $-0.154 \pm 0.033$ & $0.87 \pm 0.16$ & $0.073 \pm 0.020$ & $8.2 \pm 1.2$ & $92.3 \pm 2.1$ & $0.8 \pm 0.2$ \\ 
SNF20080913-031 & $0.0540$ & $0.087 \pm 0.042$ & $-0.013 \pm 0.041$ & $-0.015 \pm 0.041$ & $0.011 \pm 0.041$ & $-0.053 \pm 0.041$ & $-0.19 \pm 0.18$ & $0.080 \pm 0.020$ & $10.0 \pm 1.6$ & $117.8 \pm 2.4$ & $0.0 \pm 0.0$ \\ 
SNF20080914-001 & $0.0235$ & $0.498 \pm 0.020$ & $0.348 \pm 0.017$ & $0.195 \pm 0.017$ & $0.139 \pm 0.016$ & $-0.007 \pm 0.017$ & $-1.12 \pm 0.17$ & $0.205 \pm 0.024$ & $16.1 \pm 1.2$ & $129.6 \pm 3.1$ & $1.0 \pm 0.1$ \\ 
SNF20080918-000 & $0.0661$ & $0.417 \pm 0.038$ & $0.323 \pm 0.037$ & $0.173 \pm 0.037$ & $0.176 \pm 0.037$ & $0.167 \pm 0.037$ & $1.19 \pm 0.31$ & $0.176 \pm 0.021$ & $7.4 \pm 3.7$ & $152.6 \pm 5.5$ & $-0.8 \pm 0.4$ \\ 
SNF20080918-002 & $0.0857$ & $-0.286 \pm 0.048$ & $-0.086 \pm 0.047$ & $-0.014 \pm 0.047$ & $0.021 \pm 0.047$ & $0.091 \pm 0.057$ & $-1.12 \pm 0.15$ & $-0.059 \pm 0.014$ & $16.4 \pm 2.5$ & $100.8 \pm 4.5$ & $0.1 \pm 0.2$ \\ 
SNF20080918-004 & $0.0499$ & $0.223 \pm 0.221$ & $0.177 \pm 0.221$ & $0.217 \pm 0.221$ & $0.222 \pm 0.221$ & $0.232 \pm 0.221$ & $-2.07 \pm 0.30$ & $-0.042 \pm 0.026$ & $24.3 \pm 1.7$ & $90.4 \pm 8.3$ & $-0.1 \pm 0.3$ \\ 
SNF20080919-001 & $0.0409$ & $-0.526 \pm 0.028$ & $-0.467 \pm 0.027$ & $-0.482 \pm 0.028$ & $-0.422 \pm 0.027$ & $-0.446 \pm 0.028$ & $0.44 \pm 0.18$ & $0.059 \pm 0.017$ & $5.1 \pm 0.5$ & $83.5 \pm 1.2$ & $-0.2 \pm 0.1$ \\ 
SNF20080919-002 & $0.0547$ & $0.486 \pm 0.020$ & $0.583 \pm 0.017$ & $0.591 \pm 0.018$ & $0.518 \pm 0.017$ & $0.440 \pm 0.036$ & $-2.77 \pm 0.35$ & $0.064 \pm 0.034$ & $31.4 \pm 4.1$ & $107.4 \pm 22.3$ & $-2.0 \pm 0.2$ \\ 
SNF20080920-000 & $0.0390$ & $-0.161 \pm 0.078$ & $-0.196 \pm 0.077$ & $-0.241 \pm 0.077$ & $-0.180 \pm 0.077$ & $-0.174 \pm 0.077$ & $0.82 \pm 0.24$ & $0.092 \pm 0.020$ & $5.6 \pm 3.7$ & $137.9 \pm 3.3$ & $-1.3 \pm 0.3$ \\ 
PTF09dlc$^{i}$ & $0.0663$ & $-0.255 \pm 0.036$ & $-0.208 \pm 0.035$ & $-0.157 \pm 0.035$ & $-0.118 \pm 0.035$ & $-0.071 \pm 0.035$ & $-0.07 \pm 0.11$ & $-0.010 \pm 0.010$ & $12.9 \pm 1.2$ & $143.9 \pm 3.0$ & $-0.7 \pm 0.2$ \\ 
PTF09dnp$^{i}$ & $0.0376$ & $-0.491 \pm 0.060$ & $-0.242 \pm 0.059$ & $-0.138 \pm 0.059$ & $-0.227 \pm 0.059$ & $-0.369 \pm 0.059$ & $-1.62 \pm 0.29$ & $-0.024 \pm 0.023$ & $16.8 \pm 1.1$ & $65.5 \pm 8.3$ & $-1.3 \pm 0.3$ \\ 
PTF09fox$^{i}$ & $0.0706$ & $-0.248 \pm 0.033$ & $-0.185 \pm 0.033$ & $-0.074 \pm 0.033$ & $0.025 \pm 0.033$ & $0.210 \pm 0.033$ & $-0.21 \pm 0.23$ & $-0.104 \pm 0.013$ & $9.4 \pm 1.4$ & $123.8 \pm 2.7$ & $0.4 \pm 0.2$ \\ 
PTF09foz$^{i}$ & $0.0532$ & $-0.007 \pm 0.043$ & $-0.080 \pm 0.043$ & $-0.063 \pm 0.043$ & $-0.121 \pm 0.043$ & $-0.127 \pm 0.043$ & $-1.53 \pm 0.31$ & $0.053 \pm 0.023$ & $22.7 \pm 1.2$ & $127.4 \pm 2.6$ & $0.8 \pm 0.3$ \\ 
 \hline
      \multicolumn{11}{p{\textwidth}}{\small Discovery: $^{a}$Boles,
        Armstrong; $^{b}$LOSS ; $^{c}$Sloan Digital Sky Survey II;
        $^{d}$Prasad, Li (LOSS); $^{e}$Orff, Newton; $^{f}$Parisky, Li
        (LOSS); $^{g}$Monard; $^{h}$Quimby et al; $^{i}$PTF;
        . Discovery attribution was obtained from
        http://www.cfa.harvard.edu/iau/lists/Supernovae.html and the
        CBAT. $^{1}$Redshifts are taken from Cooke et al 2011 and
        Childress et al, in prep.}
    
    \end{longtable}
  \end{landscape}
} 

\end{document}